# A Biochemical Logic Approach to Biomarker-Activated Drug Release

Vera Bocharova,[†] Oleksandr Zavalov,[‡] Kevin MacVittie,[†] Mary A. Arugula,[†]
Nataliia V. Guz,[‡] Maxim E. Dokukin,[‡] Jan Halámek,[†]
Igor Sokolov,[‡] Vladimir Privman,[‡] Evgeny Katz[†]

[†] Department of Chemistry and Biomolecular Science, and
[‡] Department of Physics, Clarkson University, Potsdam, NY 13699, USA

The present study aims at integrating drug-releasing materials with signal-processing biocomputing systems. Enzymes alanine transaminase (ALT) and aspartate transaminase (AST)—biomarkers for liver injury—were logically processed by a biocatalytic cascade realizing Boolean AND gate. Citrate produced in the system was used to trigger a drug-mimicking release from alginate microspheres. In order to differentiate low vs. high concentration signals, the microspheres were coated with a protective shell composed of layer-by-layer adsorbed poly(L-lysine) and alginate. The alginate core of the microspheres was prepared from $Fe^{3+}$-cross-linked alginate loaded with rhodamine 6G dye mimicking a drug. Dye release from the core occurred only when both biomarkers, ALT and AST, appeared at their high pathophysiological concentrations jointly indicative of liver injury. The signal-triggered response was studied at the level of a single microsphere, yielding information on the dye release kinetics.

---

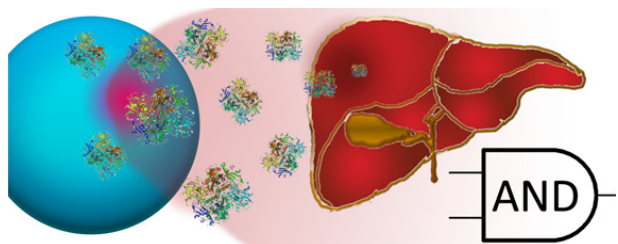

An integrated system logically processing biomarkers and releasing a drug-mimicking material was developed to demonstrate a new "Sense-Act-Treat" concept.

1.  INTRODUCTION

The design of new controlled-release chemical systems has received considerable attention due to their advantages in various areas including fragrance release,[1] self-healing materials[2,3] and particularly for delivery of bioactive substances (drugs, vitamins, nutrients, contrasts for imaging, genes, etc.) in various biomedical applications.[4-10] The therapeutic potential of these systems is especially strong because they can provide solutions for the easy control of drug transport, tackling the problem of localized delivery to a specific site.[11-13] Various functional materials based on polymer thin-films,[14-16] membranes,[17] nanoporous[18] and mesoporous[19,20] structures, capsules[21-23] and liposomes[24,25] are capable of entrapping various molecular species and nano-objects and then releasing them in response to different physical or chemical signals. A variety of polymers[9-12] have been used over the past few decades for encapsulating drugs and their subsequent delivery. Persistent challenges associated with loading of the desired molecules with high efficiency, preserving activity and stabilizing the encapsulated species, and preventing their leaching, require the development of new polymer materials and techniques to design systems with complex architectures. Among the different techniques, the polyelectrolyte layer-by-layer (LbL) self-assembly has emerged as an easily utilized and versatile tool.[26-29] Based on alternating adsorption of polycations and polyanions, this technique allows films to be produced with tunable properties; an almost infinite variety of architectures can thus be obtained.[30-32] Stimuli-responsive LbL-polymeric assemblies have been used in particular for the controlled release of drugs.[33-36]

Alginate, a natural polymer,[37] due to its appealing adhesive and mechanical properties together with availability, compatibility with hydrophobic and hydrophilic molecules, and lack of toxicity, has attracted increasing interest. Alginate hydrogels, ionically cross-linked in the presence of multivalent cations, have been extensively investigated owing to their biocompatibility.[37] Notably, alginate hydrogels have been utilized as microcapsules that can release entrapped components either spontaneously or in response to changes in environmental conditions, through controlled degradation of the assembly.[38-43] However, it is known that ionically cross-linked alginate hydrogel is not structurally stable, because of ion exchange, especially under physiological conditions, limiting its medical application.[37] A number of studies have attempted to address this problem by varying the molecular weight[44,45] or/and cross-linking ions[45] of alginate. Since alginate contains carboxylic groups on polyguluronate units, alginate microspheres exhibit negative surface charge,[46] allowing them to be used as negatively charged templates for polyelectrolyte LbL assembly. Different polyelectrolytes have been used to coat alginate,[47-53] where the most common system involves alginate/poly-L-lysine/alginate (APA) assembly,[54-57] derived from the original protocol of Lim and Sun.[58] This system consists of an alginate core ionically cross-linked with multivalent cations (often $Ca^{2+}$).[59] These microspheres are coated with poly(L-lysine) to strengthen the outer surface and control permeability,[60] followed by coating with a capping layer of adsorbed alginate to shield the cationic poly(L-lysine) from the environment and, thus, make the microcapsules biocompatible.[61]

Due to the intrinsic ability of polymers to alter their physical and chemical properties in response to external stimuli, utilization of polymers in drug delivery fosters development of systems triggered by different signals, particularly by pH changes,[62-66] temperature variation[66] and presence of specific biomolecules (e.g., glucose).[66,67] Biochemical reactions resulting in change of pH, temperature or increase/decrease in concentration of certain molecules can therefore initiate drug release. It is known that biochemically induced changes in physiological conditions often relate to the occurrence of a certain biological dysfunction or disease. Therefore design of systems triggered directly by changes in physiological conditions has been of high importance for medical diagnostics and physiologically regulated drug release. Thus far, triggered-release reactions have been accomplished via the interaction of signal-responsive materials with a single external signal, e.g., the presence of one specific kind of biomolecule.[66,67] Development of signal-responsive materials recognizing the concentration changes of several biomolecules simultaneously and responding only to specific combinations of the (bio)chemical signals would enable new biomedical applications.[68]

Information processing by biomolecules (shortly, "biocomputing"),[69,70] originally studied as a sub-area of unconventional computing,[71,72] has recently been researched as a new important tool for biosensing[73,74] and bioactuating.[75] Biomedical application of biocomputing systems allows logic analysis of complex patterns of simultaneously varying biomarkers jointly characteristic of different pathophysiological conditions.[76-78] Particularly, enzyme-based biocomputing systems[79] in the form of single logic gates[80-82] or complex logic networks[83] were utilized for the analysis of biomarkers signaling different injury conditions. Integration of enzyme logic gates with switchable electrodes allowed activation of bioelectrochemical systems by appropriate combinations of biomarkers appearing at elevated pathophysiological concentrations.[84] It should be noted that the designed switchable bioelectrodes were controlled by complex patterns of biochemical signals[85] rather than a single physical or chemical signal.

The present paper represents the next step in the development of biochemical systems controlled by logically processed biochemical signals with the result that a drug-mimicking material is released only when both biomarker-signals appeared at their high pathophysiological concentrations jointly indicative of liver injury. Thus, this work extends the research on multi-input biosensors with the built-in biomolecular logic to include an "actuator" which responds to the output signal of the binary logic processing. The signal-triggered response was studied at the level of a single microsphere, yielding information on the release kinetics of the drug-mimicking dye.

2. EXPERIMENTAL SECTION

*Chemicals and reagents.* Alanine transaminase from porcine heart (ALT, E.C. 2.6.1.2), aspartate transaminase, type 1 (AST, E.C. 2.6.1.1), citrate synthase from porcine heart (E.C. 2.3.3.1), glutamic acid, pyruvic acid, aspartate sodium salt, acetyl coenzyme A sodium salt (acetyl CoA), 5,5'-dithio*bis*(2-nitrobenzoic acid) (DTNB), alginic acid sodium salt from brown algae (medium viscosity, ≥ 2,000 cP), poly-L-lysine hydrobromide (mol. wt. 30,000-70,000), fluorescein isothiocyanate (FITC)-labeled poly-L-lysine (mol. wt. 30,000-70,000), rhodamine 6G, 3-(*N*-morpholino)propanesulfonic acid (MOPS buffer) and other standard organic / inorganic chemicals were purchased from Sigma-Aldrich, J. T. Baker and Fisher Scientific and used as supplied without any pretreatment or further purification. All solutions were prepared using ultrapure water (18.2 MΩ·cm; Barnstead NANOpure Diamond). The experiments were performed at ambient temperature, 25°C ± 2°C.

*Instrumentation and measurements.* Fluorescent measurements were performed using fluorescent spectrophotometer (Varian, Cary Eclipse). Confocal micrographs were taken with Scanning Laser Confocal Fluorescence Microscope, Nikon Eclipse C1 (488 nm argon-ion laser was used) with 10×-CFI objective. Images from the confocal microscope were handled using EZ-C1 Software Version 3.10 (Nikon Corporation).

*Fabrication of rhodamine 6G-loaded alginate hydrogel microspheres.* Sodium alginate was dissolved in water at 37ºC overnight to yield 1% (w/v) solution. Then 1.5 mL alginate solution was vigorously mixed with 0.1 mL, 0.9 mg/mL, rhodamine 6G solution in 20 mM MOPS buffer, pH 7.4. Note that the rhodamine 6G dye was used to mimic a drug molecule. The viscous alginate-rhodamine solution was pressed through a 31-gauge needle forming droplets. Compressed air (ca. 125 PSI, 30 L/min) was used to shear the droplets coming out of the tip of the needle. Alginate hydrogel formation in the form of microspheres occurred upon contact of the droplets with 1% (w/v) $FeCl_3$, dissolved in deionized water, for 15 min. After formation the microspheres were filtered from the solution using a gravity funnel apparatus and transferred to a Petri dish where they were washed with water twice and stored in deionized water at room temperature.

*LbL assembly of poly-L-lysine /alginate on the alginate microspheres.* Poly-L-lysine and alginate were dissolved in concentrations of 0.05% (w/v) and 0.5% (w/v) in 20 mM MOPS buffer, pH 7.4, and deionized water respectively. These solutions were used to deposit first a layer of poly-L-lysine and then a second layer of alginate on the $Fe^{3+}$-cross-linked alginate microspheres loaded with rhodamine 6G. Each deposition step proceeded for 5 min and the coated microspheres were washed with water after each polymer adsorption step. The final alginate/poly-L-lysine/alginate (APA) microspheres were composed of a central core consisting of the $Fe^{3+}$-cross-linked alginate hydrogel loaded with rhodamine 6G dye coated with electrostatically bound polymer shells of poly-L-lysine and sodium alginate.

*The characterization of the alginate hydrogel microspheres.* The size distribution of the rhodamine 6G-loaded microspheres in aqueous solutions before drug-mimicking release was visualized by confocal microscopy. The rhodamine 6G dye was excited with a 488 nm argon laser and became visible using scanning laser confocal fluorescence microscopy under 10×-objective lens. The produced microspheres ranged in size from 500 to 800 μm, however, only microspheres with the size of ca. 500 μm were selected for the experiments. FITC-labeled poly-L-lysine was used to visualize the poly-L-lysine layer applied to the microspheres. The microspheres treated with FITC-labeled poly-L-lysine were produced in the same way as described above, however the addition of rhodamine 6G dye to the initial sodium alginate solution was skipped because the FITC-label has excitation and emission spectrum peaks (495 nm and 521 nm respectively) similar to those of rhodamine 6G.

*The composition of the biocatalytical cascade for the liver injury analysis.* Glutamic acid (10 mM), pyruvic acid (5 mM), aspartate sodium salt (50 mM), acetyl CoA (5 mM), and citrate synthase (1 U/mL) were dissolved in phosphate buffer saline (PBS) (137 mM NaCl, 2.7 mM KCl, 10 mM $Na_2HPO_4$, 2 mM $KH_2PO_4$), pH = 7.4, to form the "machinery" of the system. Logical '**0**' and '**1**' levels of ALT (0.02 and 2 U/mL), and AST (0.02 and 2 U/mL) which correspond to the normal and pathophysiological biomarker concentrations,[86] respectively, were used as inputs *A* and *B* activating the biochemical "machinery" system. This system works as an **AND** gate, signaling presence of the liver injury when both inputs appear at their logic **1** concentrations. Production of citrate was indirectly measured for each combination of the inputs using interaction between DTNB[87,88] reagent and a thiol group of CoASH which is produced by the biocatalytic cascade in 1:1 stoichiometry to citrate, see Scheme 1 (all the Schemes and Figures are located at the end of this preprint, starting on page 25). For this analysis, after 1 hour of the biocatalytic reaction, 50 μL aliquot of the reaction mixture was reacted with 10 μL of DTNB (0.1 mM) and the absorbance at 412 nm was used to calculate[87,88] the concentration of the produced CoASH. The concentration of citrate was assumed to be the same. In order to determine the total amount of the loaded rhodamine 6G dye, individual microspheres were fully dissolved in 10 mM citrate solution for 30 min and the amount of the released rhodamine 6G was found by fluorescence measurements.

*In vitro release study.* In order to measure the rhodamine 6G dye release to the solution, triggered by the biocatalytic cascade for the analysis of liver injury, single microspheres were isolated and placed in different cuvettes containing the "machinery" system. By application of different ALT/AST input combinations to different cuvettes the fluorescence of the released rhodamine 6G was measured every 5 min for the system with mixing and every 10 min for undisturbed system. Mixing was done with pipette 3 times up and down each time before the measurement. For the experiments without mixing, a light beam passed through a horizontal slit 1 mm wide to excite fluorescence at a thin slice of solution inside the cuvette at 6 mm above the bottom, where a single rhodamine 6G-loaded microsphere was placed.

In order to directly visualize the kinetics of the microsphere dissolution, a single microsphere was placed on a microscope glass slide and observed by using a confocal microscope. The microspheres were dissolved in different solutions mimicking production of citrate *in situ* from the biocatalytic cascade. Citrate, 0.34 mM solution, was used to mimic the result of the application of ALT/AST inputs in **1,0** combination, while 1 mM solution was selected to mimic the **1,1** input combination. In a control experiment ascorbic acid, 1 mM, was used to visualize the kinetics of the microsphere dissolution and compare it with the equivalent concentration of citrate. The kinetics of the dissolution was represented as change in pixels brightness of the background which was found by demarcating a small square of pixels on the image and monitoring their brightness over time, as the microspheres dissolved. The analyzed square was drawn in the top-left corner of each image.

Movies showing time-dependent dissolution of the APA microspheres in the presence of 0.34 and 1 mM citrate are available in the Electronic Supplementary Information. The imaging was performed by using Scanning Laser Confocal Fluorescence Microscope, Nikon Eclipse C1 (488 nm argon-ion laser was used with 10×-objective) where pictures were taken every 2 min. Movies were composed in Windows Movie Maker program from the pictures obtained from the microscope. The movie for 0.34 mM of citrate, composed of 30 frames, corresponds to the monitoring of the microsphere dissolution during the time of 60 min. The movie for 1 mM of citrate, composed of 13 frames, corresponds to the observation of the microsphere dissolution during 26 min.

## 3. RESULTS AND DISCUSSION

Scheme 1 shows the general composition and operation of the system with the drug-mimicking release from the alginate microspheres triggered by a biocatalytic cascade activated by biomarkers signaling liver (hepatic) injury. This system is based on the recently developed concept of two non-specific biomarkers operating together, and it offers a high fidelity indication of liver injury when both biomarkers appear at their elevated pathophysiological concentrations.[73,74,80-82] The biomarker selection in each case is based on their biomedical relevance to the specific pathophysiological condition and on their ability to operate jointly in a biocatalytic cascade producing the final desired product. In order to provide a binary "YES/NO" conclusion on specific pathophysiological conditions, for an analytical application the end product generated by the biocatalytic cascade should be easily analyzed by optical[80-82] or electrochemical[81] means. The present work is focused on the actuation involved in the model drug release, and therefore the final product generated by the biocatalytic cascade should dissolve the alginate microcapsules, releasing the loaded dye.

Two enzymes, alanine transaminase (ALT) and aspartate transaminase (AST), were selected as biomarkers signaling liver injury. While they appear in human blood at concentration of 0.02 U/mL under normal physiological conditions, their elevated concentrations of 2 U/mL correspond to pathophysiological conditions characteristic of the liver injury.[86] Therefore, we defined 0.02 and 2 U/mL concentrations as logic inputs **0** and **1**, respectively, while the "gate machinery" included the rest of the system, see Scheme 1, identical for each combination of the logic inputs. Since all experiments were performed *in vitro*, the "machinery" solution was activated with the model solutions containing the ALT and AST biomarkers at the concentrations of 0.02 or 2 U/mL, mimicking the *in vivo* physiological conditions. The biomarker inputs were added systematically in four different combinations: **0**,**0**, **0**,**1**, **1**,**0** and **1**,**1** (for ALT,AST, respectively) where **0**,**0** and **1**,**1** combinations corresponded to the normal physiological and pathophysiological conditions of liver injury, respectively, while combinations **0**,**1** and **1**,**0** were not definitive, since they might result from other than liver injury biological dysfunctions. Two reactions biocatalyzed by ALT and AST were coupled in the biocatalytic cascade through α-ketoglutarate, Scheme 1. Even though logic **0** inputs were not the physical zero concentrations of the enzymes, and thus the biocatalytic cascade was active for all four combinations of the input signals, only **1**,**1** combination of the inputs resulted in higher rates of both biocatalytic reactions, yielding higher concentration of oxaloacetate.

Oxaloacetate generated by the joint activity of the two input enzymes was converted to citrate with the help of the third enzyme, citrate synthase. This reaction fostered generation of the species capable of dissolving the alginate microspheres and releasing a drug-mimicking dye. It is worth mentioning that the byproduct of the third biocatalytic reaction, CoASH, was analyzed through its reaction with DTNB, Scheme 1. Since CoASH was produced in the 1:1 stoichiometry to citrate, the concentration of citrate was determined on the basis of production of CoASH in the presence of different combinations of the enzyme-input signals, Scheme 1, inset. Application of the **1**,**1** input combination corresponding to the liver injury resulted in 1 mM citrate after 60 min reaction, while the highest citrate concentration corresponding to the conditions different from liver injury was ca. 3-fold lower (0.34 mM) and it was achieved at the **1**,**0** input combination. After 1 h the activity of enzymes comes to saturation, so after that time increase in citrate concentration was not observed. Aiming at the drug-mimicking release triggered specifically by the **1,1** input combination corresponding to the liver injury, the kinetics of dissolution of the alginate microspheres should be significantly different in the presence of 0.34 mM and 1 mM citrate, as demonstrated in the following sections of this paper.

Recently, we reported an electrochemical fabrication of alginate thin-films cross-linked with $Fe^{3+}$ cations.[89] In the present study we use the ability of the $Fe^{3+}$ cations to form a complex with citrate: binding constant $K_1$ ($logK_1$ = 11.85).[90] Thus, a ligand exchange reaction between citrate and alginate in the $Fe^{3+}$-alginate complex initiates the alginate hydrogel dissolution. Note that the $Ca^{2+}$ cations frequently used for the alginate cross-linking have a much weaker complex formation with citrate ($logK_1$ = 3.5).[90] Therefore, citrate is not able to dissolve alginate hydrogels if they are cross-linked with $Ca^{2+}$ cations. It should be noted that the present study aims at selective dissolution of the alginate microspheres at the citrate concentration of 1 mM achieved in the presence of **1,1** combination of ALT/AST biomarkers, while the lower concentrations of citrate (up to 0.34 mM at **1,0** input combination) should not result in the microsphere dissolution. Thus, the microspheres should be capped with protecting layers discriminating the effects of citrate at 0.34 and 1 mM concentrations. This was achieved by LbL deposition of poly-L-lysine and alginate resulting in the alginate/poly-L-lysine/alginate (APA) microspheres, Scheme 2. The APA microspheres were visualized with the confocal microscope, Figure 1 (all the Figures are located at the end of this preprint, starting on page 27). While using FITC-labeled poly-L-lysine, the thickness of the inner poly-L-lysine layer was estimated as ca. 18 μm and the size of $Fe^{3+}$-cross-linked alginate core loaded with the rhodamine 6G dye was ca. 500 μm.

The APA microspheres have demonstrated significant difference in their dissolution when they were reacted with 0.34 and 1 mM citrate solutions. Figure 2 depicts selected images of a single APA microsphere dissolution, obtained with the confocal microscope. Figure 3 shows the dynamics of the dissolution derived from the images obtained with the confocal microscopy. The APA microsphere was rapidly dissolved after the lag-period of ca. 7 minutes when it was reacted with 1 mM citrate solution. On the other hand, the APA microsphere was reacting very slowly with 0.34 mM citrate over the period of time longer than 60 min. (Movies showing time-dependent dissolution of the APA microspheres in the presence of 0.34 and 1 mM citrate are available in the Electronic Supplementary Information, which is **available at the journal web site**). In a control experiment an APA microsphere was reacted with 1 mM ascorbic acid which is known to react with $Fe^{3+}$-cross-linked alginate hydrogel, reducing the $Fe^{3+}$ cations and thus dissolving the hydrogel. Importantly, the protecting layers in the APA structure also slow down the microsphere dissolution in the presence of ascorbic acid preventing its penetration into the $Fe^{3+}$-cross-linked alginate core. This result is particularly important for potential future applications of the APA microspheres *in vivo* where ascorbic acid that is present in blood (ca. 1 mM) and might be one of the main interferants. We note that citrate is also present in blood,[91] but at concentrations (0.12 mM) which are too low to initiate the dissolution of the alginate microspheres.

After studying the dissolution of a single APA microsphere, we investigated the process of the release of the drug-mimicking dye into the solution. In the first set of experiments the release was performed under mixing of the solution, thus enhancing the mass-transport of the dye after the dissolution of the APA microsphere induced by different combinations of the ALT/AST input signals. Figure 4 shows the kinetics of the fluorescence increase obtained from a single APA microsphere when introduced into a solution containing **0**,**0**, **0**,**1**, **1**,**0** or **1**,**1** combinations of the input signals. Notably the input combinations **0**,**0**, **0**,**1** and **1**,**0** result in a very slow and small increase of the fluorescence, thus proving that the protecting two-layer shell on the APA microsphere keeps the dye-loaded core intact. On the other hand, the input combination **1**,**1** results in a fast significant increase of the fluorescence after the lag-period of ca. 40 min. In the control experiment performed in the presence of all components of the biocatalytic cascade except citrate synthase, thus prohibiting citrate formation, the dissolution of the microspheres was not observed within the studied time scale.

Additional information on the properties of the APA microspheres, specifically, the process of release of the dye from the microspheres, can be obtained in experiments whereby the dye diffuses without any initial mixing, upwards along the cuvette from its bottom where a single microsphere is located. Figure 5 shows the data sets taken for various concentrations of citrate. A standard diffusion theory was used for the time-dependence of the cross-section-averaged concentration at height, $x$, upwards along the cuvette, similarly to the method used in the previous work.[92] The concentration, $c(x,t)$, averaged over the cuvette cross-section, was measured by fluorescent spectroscopy at $x = $ 6 mm, as described in the experimental section. The diffusion equation assuming that the microsphere size is small as compared to other dimensions, was taken as

$$\frac{\partial c}{\partial t} = D \frac{\partial^2 c}{\partial x^2} + S(t)\delta(x), \qquad (1)$$

for $0 \leq x < \infty$, with the reflecting boundary condition at $x = 0$, and with the source term, $S(t)\delta(x)$, determined by the dye release. Here $\delta(x)$ is the Dirac delta-function representing a point source, whereas the expression for $S(t)$ is given in the next paragraph. The diffusion constant of the dye molecules, $D = 2.8 \times 10^{-10}$ m$^2$sec$^{-1}$, is known.[93]

We assume that the time scale of dissolution of the shell is $t_0$, after which the core of radius $R_0$ remains, containing $N$ molecules of dye. As this core is decomposed, or the dye is leached out from it, the radius of the remaining dye-containing spherical volume, $R(t)$, will decrease to zero. This process involves another time scale, $t_c$. The quality of the available data only allows extraction of information on these two parameters, and therefore a relatively simple model of the dye etching should be used. For the present system, the concentration of the dye in the solution is too low, and its transport away from the dissolving core too fast to affect the rate of dye release. Basically, this corresponds to the assumption that in the Noyes–Whitney law,[94] the concentration gradient in the near-surface diffusion-controlled layer is not significantly time- or core-size dependent, and that the external (to that layer) concentration is small as compared to the internal one (at the dissolving core surface). Then the dye is released at a constant rate per unit area of the core of radius $R(t)$, and therefore the volume, $V = 4\pi R^3/3$, of the dye-containing core shrinks according to $\frac{\partial V}{\partial t} \propto 4\pi R^2$. One can then easily show that this approach leads to a linear decrease $R(t) = R_0 \left(1 - \frac{t-t_0}{t_c}\right)$. Thus, the core radius decreases from $R_0$ at $t_0$ to zero at time $t_0 + t_c$. The amount of dye released can then be calculated, and the source term in the diffusion equation turns out to be given by $S(t) = \frac{3N}{AR_0^2 t_c} R^2(t)$ for times $t_0 < t < t_0 + t_c$, and $S(t) = 0$ otherwise, where $A$ is the cross-section area of the cuvette.

The diffusion equation can now be solved analytically for $0 \leq x < \infty$, where no upper limit is assumed because it can be estimated that the height of the column of solution in the cuvette, 12 mm, will affect the concentration only after times of approximately 700 min, safely outside our measurement range (we also tested the latter assumption by numerical solutions with the reflecting boundary condition at the top of the liquid column). For $t_0 \leq t \leq t_0 + t_c$, we get

$$c(x,t) = \int_{t_0}^{t} d\tau \frac{3N}{At_c} \left(1 - \frac{\tau-t_0}{t_c}\right)^2 \frac{1}{\sqrt{\pi D(t-\tau)}} e^{-\frac{x^2}{4D(t-\tau)}}, \qquad (2)$$

with the upper limit of integration replaced by $t_0 + t_c$ for $t > t_0 + t_c$, and obviously $c(x,t) = 0$ for $t < t_0$. Least-squares data-fitting results, using the standard approaches (including error estimates),[95] are presented in Table 1 (see page 24). The sums of the times, $t_0 + t_c$, are generally consistent with the earlier estimates for the overall time of the dye concentration buildup measured by the microsphere dissolution in the system with mixing.

The diffusion-theory study clearly confirms that *two separate time scales* are involved in the process. This is illustrated in Figure 6, in which we used experimental parameters for all the quantities but the time scales. We note that the time scale $t_0$ simply represents the offset in the concentration buildup to positive times, due to the need to first dissolve the shell. A single additional time scale for the dye-containing core decomposition, here illustrated by varying $t_c$ values, cannot modify this offset. Rather, increasing it leads to slower concentration buildup at the distance *x* from the cuvette bottom. Note that for larger times than those of our experiment, the concentration at *x* will actually decrease, to zero if there is no reflecting liquid-column top — as assumed for Equation (2). Our time-dependence form for the remaining dye-containing core radius is sketched in the inset in Figure 6. In reality, the effective radius $R_{eff}$ defined, for instance, via $R_{eff}(t)/R_0 = (n(t)/N)^3$, were $n(t)$ is the remaining undissolved dye at time *t*, with $n(0) = N$, is likely a smooth function of time, but requiring two time scales for describing its time dependence: one for the shall and the other for the dye-containing core.

## 4. CONCLUSIONS

The present study has demonstrated that the drug-mimicking release process can be triggered by at least two biomarker-signals logically processed by a biocatalytic cascade featuring binary **AND** logic operation. Simultaneous appearance of two biomarkers at pathophysiological elevated concentrations (the **1,1** input combination) resulted in the alginate microsphere dissolution and release of the loaded dye that mimics a drug. Importantly, normal physiological concentrations of the biomarkers (the **0,0** input combination) or even a single-biomarker elevated concentration (the **0,1** and **1,0** input combinations) do not result in the "drug" release at least on the studied time-scale. The effective discrimination between the **1,1** inputs indicative of the liver injury conditions and all the other input combinations, corresponding to the normal physiology (**0,0**) or to another pathophysiology different from the liver injury (**0,1** and **1,0**), was achieved by protecting the dye-loaded alginate core with the two-layer polymer shell. Theoretical model of the dye release and transport has allowed evaluation of the time scales of the microsphere dissolution process, and discrimination between the initial shell decomposition and later release of the dye from the alginate core. Once the kinetics of the dye release from a microsphere is known based on studies in simple diffusion situations, such as reported here, the obtained information can be used as input for evaluation of dye transport in more complicated situations, involving convective diffusion and complicated flow patterns, for instance, in blood vessels, by using standard computational fluid-dynamics techniques.

Despite the fact that the studied system aims only at the demonstration of the concept, practical applications are feasible. However, after the successful implementation of the **AND** logic operation to the drug-mimicking release process still there are unsolved problems which require additional work. The present study was performed using solutions artificially spiked with the biomarker-inputs mimicking normal physiological and pathophysiological conditions (defined as logic **0** and **1** inputs, respectively). In the continued study the inputs should be natural biologically produced materials obtained from animal or human biomedical samples, similarly to our recent study where porcine samples were applied for the logic analysis of liver injury biomarkers.[99] It has been already shown that similar binary logic systems processing injury biomarkers can operate in human serum solutions.[80,96] Preliminary experiments also demonstrated that $Fe^{3+}$-crosslinked alginate can be stable in human serum and its dissolution can be triggered by chemical signals. Still additional work will be required to optimize the structure and composition of the drug releasing microcapsules for their robust operation in human serum solutions or other physiological liquids. The issue of the potential toxicity of $Fe^{2+}$ ions[97] released upon dissolution of the $Fe^{3+}$-crosslinked alginate microspheres should be addressed. The concomitant release of $Fe^{2+}$ ions upon complete decomposition of a single microsphere results in dissolution of ca. 10 μg iron ions. This amount should be compared with iron content (10-18 mg daily) in food supplements according to the Recommended Dietary Allowance (RDA) of nutritional elements.[98] One can conclude that the iron release from the microsphere is much below the recommended iron consumption, thus there is no any toxicity issue associated with the $Fe^{2+}$ ions. The last but not the least, the present study utilized a drug-mimicking material, thus limiting the practical applicability of the developed system. It should be noted that another study performed in our lab has demonstrated that the $Fe^{3+}$-crosslinked alginate thin-films can be applied for encapsulation and triggered release of lysozyme, which is a real antimicrobial drug.[100] In the next step of the project, which is underway, real drug compensating liver injury problems will be used instead of a drug-mimicking dye. This approach will also be extended to other biomarkers signaling different kinds of pathophysiological problems, and the alginate microspheres will be loaded with the corresponding drugs.

**ACKNOWLEDGMENTS**

This work was supported by the NSF (Award # CBET-1066531).

**ELECTRONIC SUPPLEMENTARY INFORMATION**

Movies showing time-dependent dissolution of the APA microspheres in the presence of 0.34 and 1 mM citrate. See DOI 10.1039/C2JM32966B.

**Table 1**: Results of fitting the time-dependent data presented in Figure 5 according to Equation (2). Note that all the parameters but the fitted values of $t_0$ and $t_c$ are known independently.

| Input combinations | $t_0$ / min | $t_F$ /min |
|---|---|---|
| **1,1** | 8.2 ± 0.6 | 49.6 ± 2.5 |
| **1,0** | 64.8 ± 3.1 | 116.4 ± 3.8 |
| **0,1** | 78.3 ± 3.5 | 123.4 ± 4.2 |

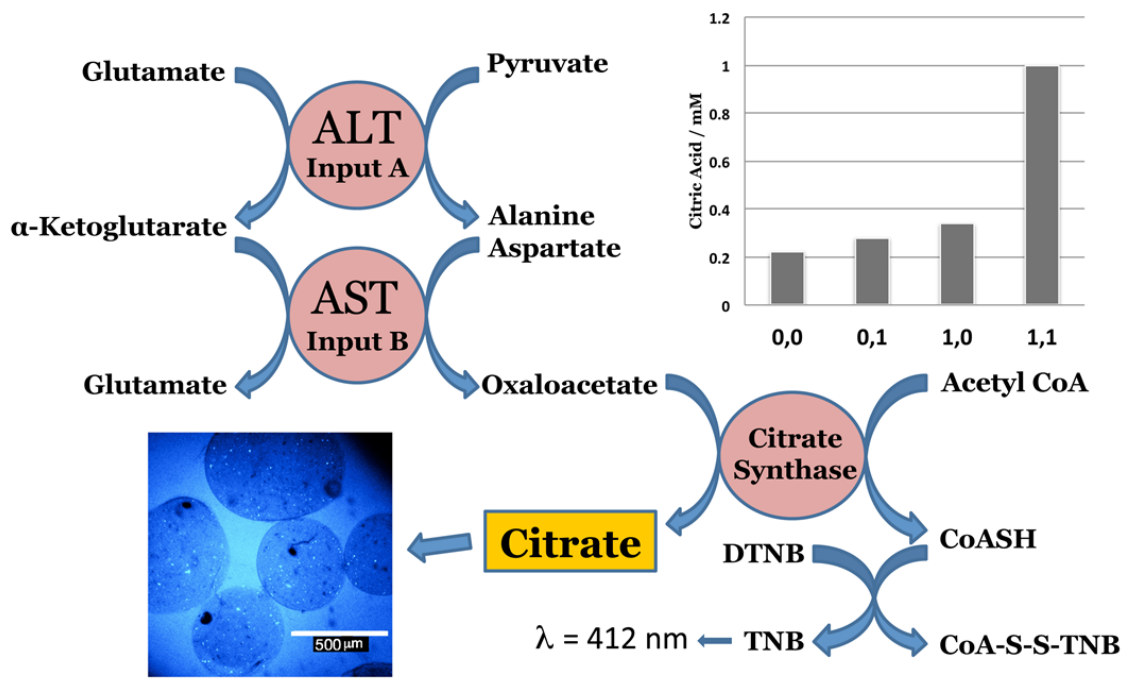

**Scheme 1.** The biocatalytic cascade mimicking the **AND** logic gate activated by two input signals, ALT and AST, which are biomarkers of liver injury. The production of citrate, which is the final product of the biocatalytic cascade, is indirectly analyzed by measuring CoASH byproduct using DTNB-assay for thiol groups. Citrate induces dissolution of the alginate microspheres shown in the micrograph.

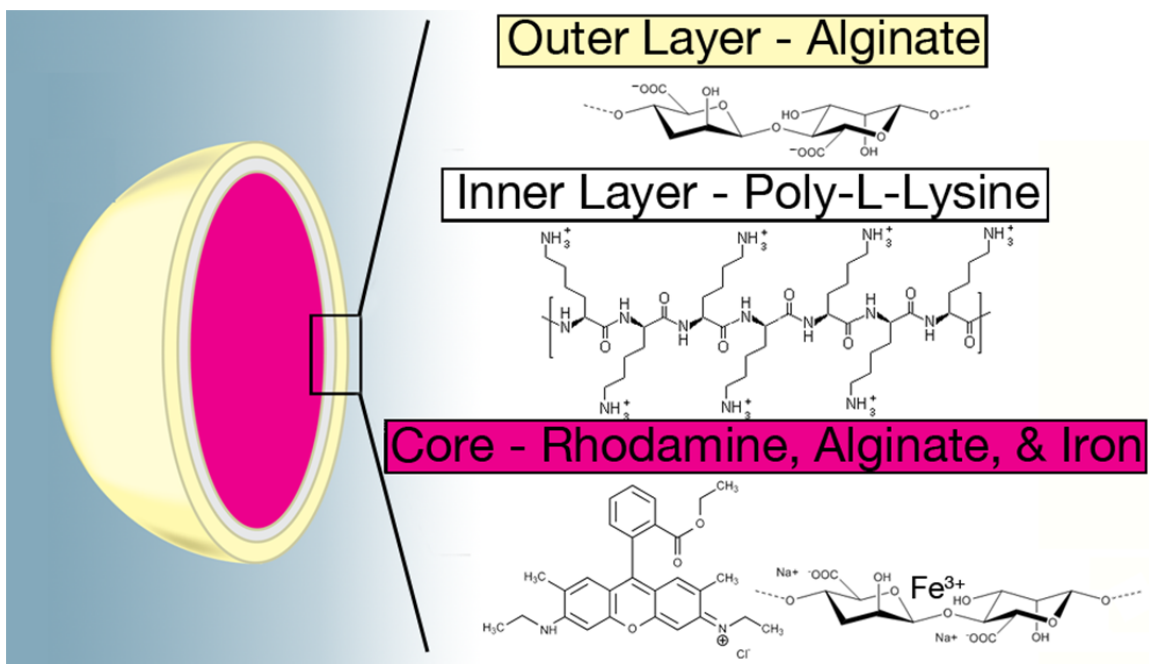

**Scheme 2.** The cartoon shows the APA microsphere composition which includes a central core of $Fe^{3+}$-cross-linked alginate gel loaded with rhodamine 6G dye and a two-layer shell consisting of a poly-L-lysine inner layer and sodium alginate outer layer.

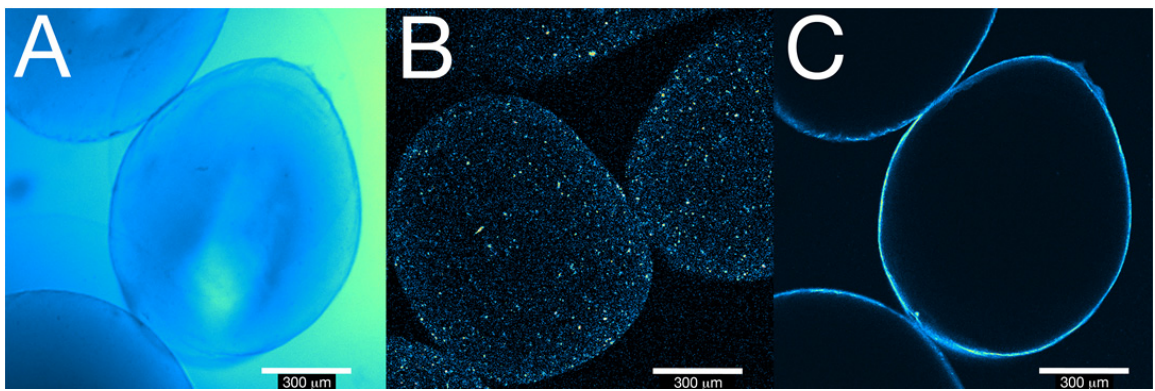

**Figure 1.** Images taken using a confocal optical microscope, where (A) is optical visualization of the microspheres (without fluorescence), (B) is a fluorescent image of the rhodamine-loaded core, and (C) shows a fluorescent image of the FITC-labeled poly-L-lysine outer layer in the shell (note that rhodamine was not included in the core).

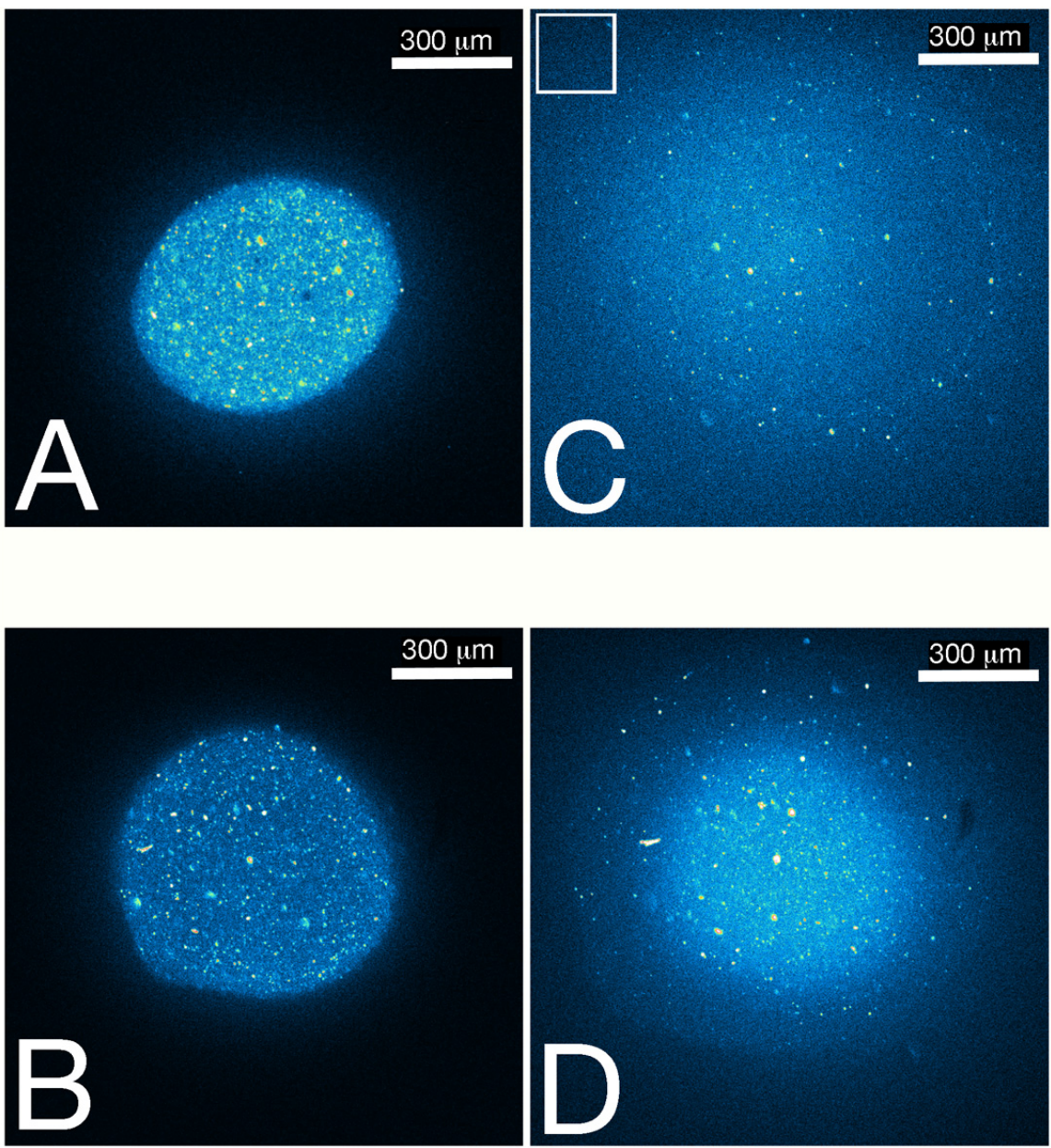

**Figure 2.** Fluorescence images of the dissolving microspheres upon reacting with the different citrate concentration and obtained with a confocal microscope after different reaction time: (A) 1 mM, 3 min, (B) 0.34 mM, 3 min, (C) 1 mM, 20 min, (D) 0.34 mM, 20 min. The progress in the microsphere dissolution can be observed when comparing images A and C, B and D. The images represent selected frames from movies which are available in the Electronic Supplementary Information.

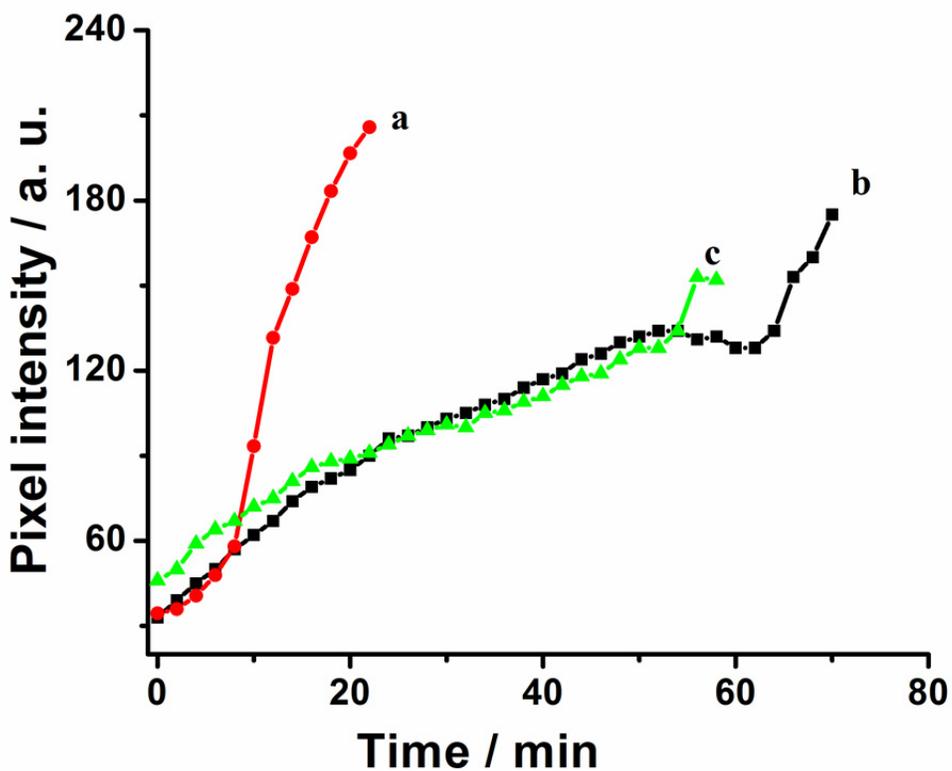

**Figure 3.** The kinetics of the microsphere dissolution upon reacting with: a) 1 mM citrate, b) 0.34 mM citrate, c) 1 mM ascorbic acid—a major possible interferant in blood. The kinetic was derived from the fluorescence images similar to those shown in Figure 2 and it was measured as the pixel brightness in a square marked in image (C) and in other similar images, Figure 2. The same image area was analyzed for brightness in all images recorded over the reaction time.

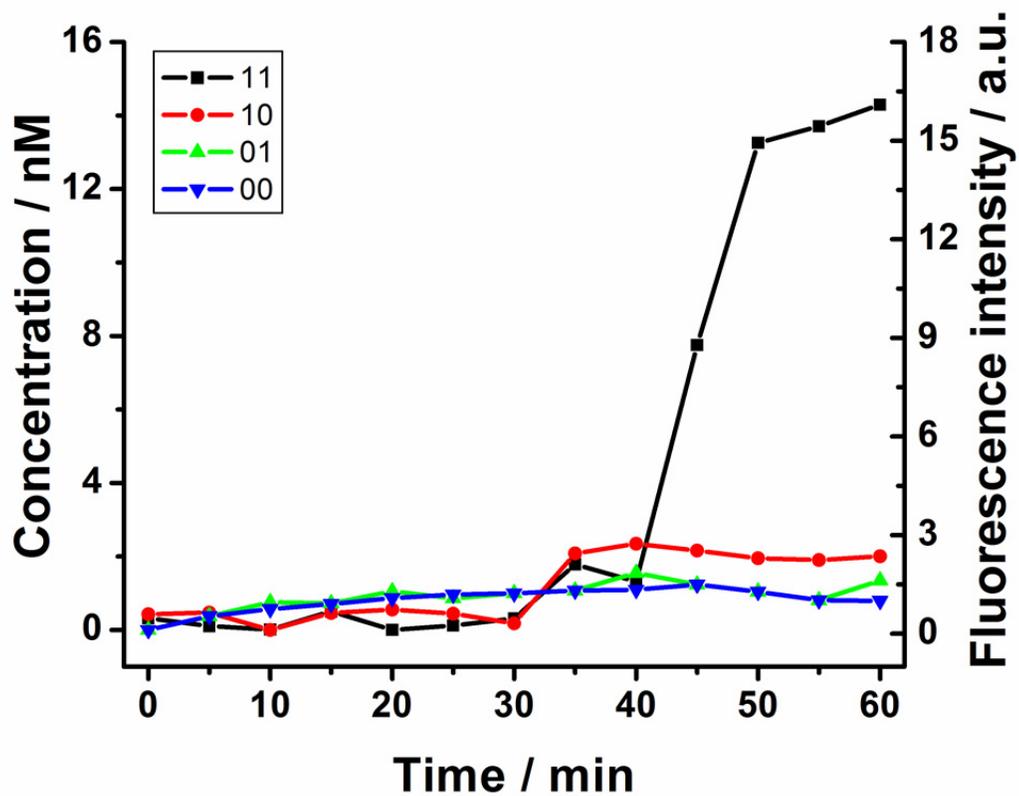

**Figure 4.** The kinetics of the rhodamine 6G release from a single APA microsphere in the presence of the biocatalytic cascade shown in Scheme 1, activated with different combinations of the ALT/AST input signals: **0**,**0**, **1**,**0**, **0**,**1** and **1**,**1**. The release was performed under mixing conditions. The experiments were performed in PBS buffer. The rhodamine 6G concentration was calculated from the fluorescence measurements using a calibration plot.

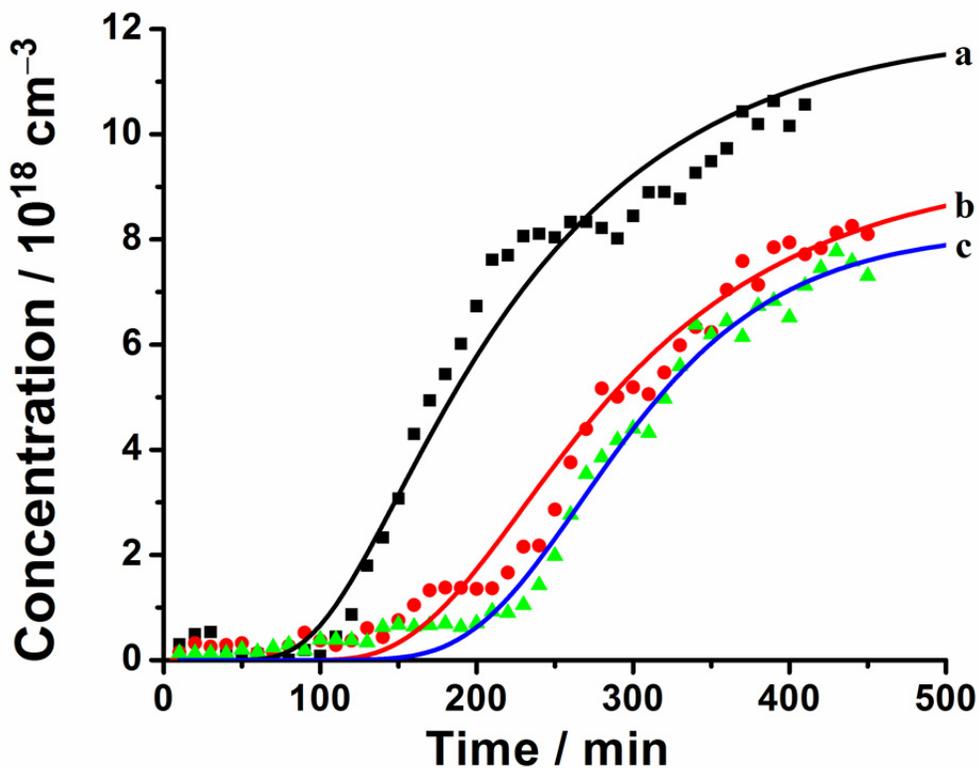

**Figure 5**.  Data sets for concentration measurements at height $x$ = 6 mm upwards along the cuvette, for various input combination a) **1**,**1**, b) **1**,**0**, c) **0**,**1**, hence different concentrations of citrate. The experiments were performed in PBS buffer. The model fitting curves are based on the diffusion theory and assumptions of the dye release kinetics as described in the text.

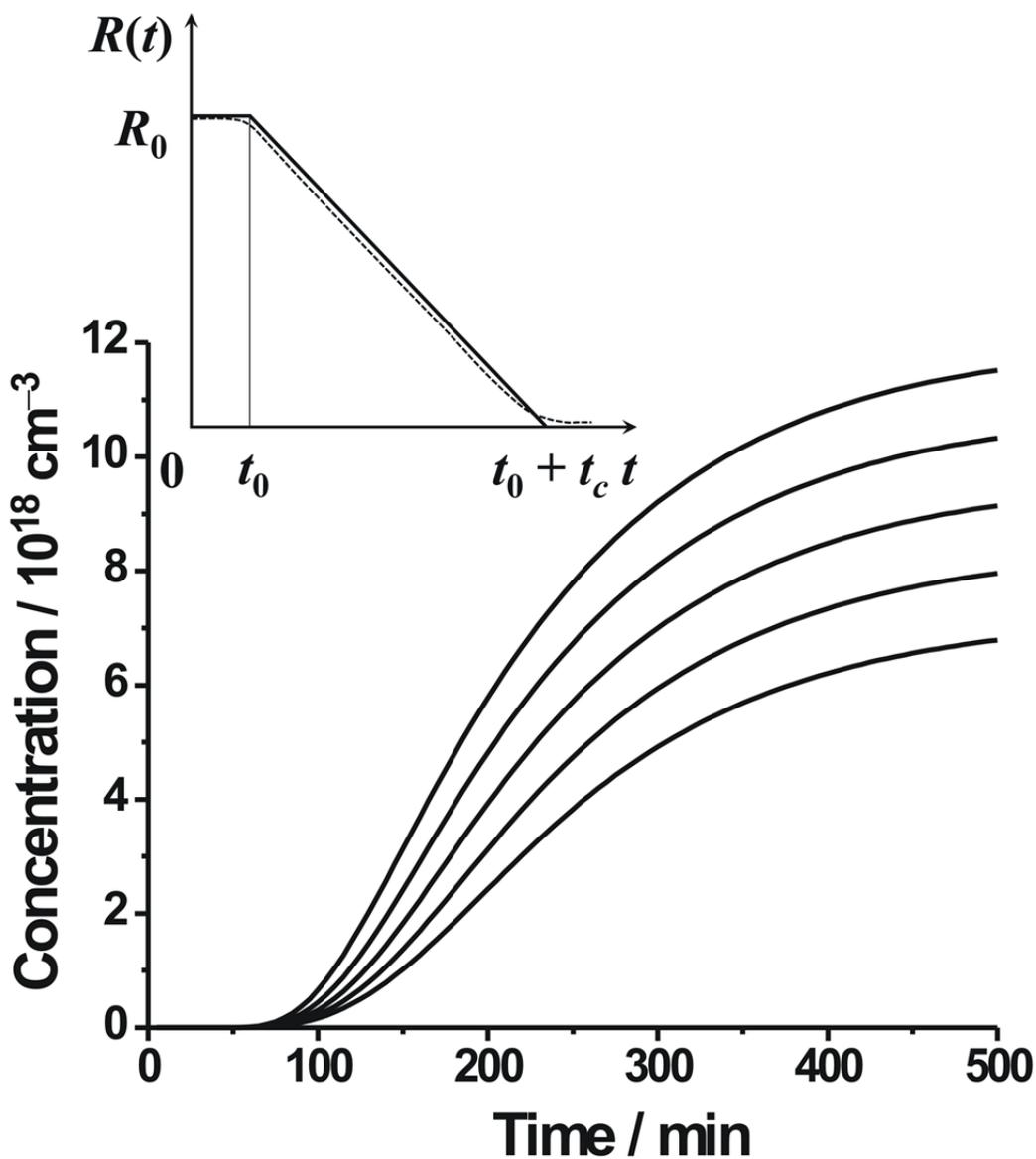

**Figure 6**: Time dependence, Equation (2), at fixed $x$, for illustrative values $t_0 = 10$ min, and, in decreasing curve heights order, $t_c = 50, 75, 100, 125, 150$ min. Note that for larger times these curves are not monotonically increasing (see text). The inset schematically shows the assumed time dependence of the dye-containing core radius (solid line) and the actual effective radius (dotted line).